\documentclass[preprint,12pt]{elsarticle}
\usepackage[english]{babel}
\usepackage[T1]{fontenc}
\usepackage[utf8]{inputenc}
\usepackage{graphicx,epstopdf}
\usepackage{amssymb}
\usepackage{amsmath}

\usepackage{amsfonts}
\usepackage{bbm}
\usepackage{xcolor}
\usepackage{latexsym}
\usepackage{color,soul}
\usepackage{listings}
\usepackage{bbold}
\usepackage{comment}
\usepackage{mathrsfs}
\usepackage[colorlinks=true, linkcolor=blue, citecolor=blue, urlcolor=blue]{hyperref}

\newcommand{\1}{\mathbbm{1}}

\begin{document}
\title{Electronic Effects of a Twisted Graphene Catenoid Bridge}

\author[1]{G. M. Delgado\corref{cor1}}
\ead{gabrieldelgadomelo@alu.ufc.br}

\author[1]{J. E. G. Silva}
\ead{euclides@fisica.ufc.br}

\cortext[cor1]{Corresponding author}

\address[1]{Universidade Federal do Cear\'a, Departamento de F\'{i}sica, 60455-760, Fortaleza, CE, Brazil}

\date{\today}

\begin{abstract}
The interplay between non-trivial background geometries and quantum dynamics has emerged as a powerful tool to tailor the electronic properties of 2D materials. In this work, we investigate the effective quantum dynamics of massless Dirac fermions confined to a twisted graphene structure called a catenoid bridge, which connects two single-layer sheets.
By adopting a continuum approach, where the electron dynamics is governed by a purely covariant curved Dirac equation, we obtain the effective Hamiltonian containing both curvature and twist interactions. We found that the torsion modifies the electronic states by producing a geometric phase on the wave function. In addition, the twist also deforms the surface geometry, which leads to a new geometric term in the effective Hamiltonian. This twist potential enhances the barrier around the catenoid throat, which increases the suppression of the inter-layer transmission coefficient. 
Like the spin-curvature interaction, the spin-twist term has a chiral dependence which is invariant under a combined parity and spin flip transformation. As a result, the electronic states can be restricted to the upper or lower layer.
These findings provide valuable insights into how mechanical deformations can be harnessed to control quantum transport in graphene-based wormhole architectures.
\end{abstract}


\maketitle

\section{Introduction}
\label{sec: Introduction}

In recent years, two-dimensional (2D) materials such as graphene~\cite{novoselov2012roadmap,geim2009graphene,geim2007rise} and silicene~\cite{kara2012review,chen} have attracted significant interest, largely because their low-energy electronic excitations behave as massless Dirac fermions, allowing for the study of emergent relativistic effects in tabletop systems. The influence of geometry on the quantum dynamics of such confined structures has become a central theme in modern condensed matter physics~\cite{bowick2009two,wang2017geometric,WANG201668,pham20222d}. A noteworthy feature of these systems is that mechanical deformations, given by curvature~\cite{POPOV200461} and torsion, can be used to decisively modulate their transport characteristics~\cite{Santos_2016}. From a quantum mechanical standpoint, deformations introduced by curvature and torsion act as effective gauge fields or geometry-induced potentials, enabling the control of fermion dynamics by fundamentally modifying the Dirac equation on the deformed surface. This concept, often termed strain engineering, has been demonstrated experimentally—for instance, through the observation of conductance oscillations in mechanically twisted nanotubes~\cite{cohen2006torsional, pantano2013electronic}—an effect that aligns with the emerging field of strain-twistronics \cite{hou2025strain}. Recent investigations have also actively explored the behavior of twisted bilayer graphene under the influence of external fields \cite{sinha2024electronic}.

In crystal lattices, topological defects such as dislocations and disclinations fundamentally alter the background geometry \cite{RevModPhys.80.61}. To describe quantum particles on such curved surfaces, various theoretical frameworks have been developed, including Dirac's constrained quantization, the geometric momentum approach \cite{LIU2017123, wang2017geometric}, and the confining formalism introduced by da Costa \cite{da1981quantum}. The latter effectively constrains a particle to a surface by introducing a geometric potential that depends solely on the local mean and Gaussian curvatures, an approach successfully applied to numerous geometries \cite{ferrari,catenoid,gomes2020electronic,delgado2026non}. While da Costa's method was originally formulated for the non-relativistic Schrödinger equation, investigating Dirac materials like graphene requires coupling the relativistic Dirac equation directly to the curved background, typically by employing the tetrad formalism.

The Dirac equation in curved spacetime \cite{collas2019dirac} generalizes its usual Minkowski counterpart to arbitrary manifolds. This generalization relies on the tetrad (or \textit{vielbein}) formalism, which establishes a local inertial frame at each point on the surface. This formulation is essential to describe materials like graphene, where the electrons behave near the Dirac points as massless relativistic Dirac fermions \cite{katsnelson2020graphene}. Within this geometric framework, the spin connection $\Omega_\mu(x)$ couples directly to the Dirac spinor, effectively acting as a geometry-induced pseudomagnetic gauge field. Consequently, the curved Dirac equation has become a standard tool to investigate the electronic properties of various deformed graphene geometries, such as Gaussian bumps \cite{de2007charge,almeida2026dirac}, the Möbius strip \cite{mobius}, and the helicoid \cite{watanabe2015electronic}.

In this paper, we investigate the effective quantum dynamics of massless Dirac fermions confined to twisted surfaces of revolution. Employing the purely covariant \textit{curved Dirac equation}, we first apply this framework to arbitrary twisted surfaces of revolution with a variable radial profile $R(z)$; we derive a one-dimensional (1D) effective equation and show that a geometric phase arises as a robust topological feature. As a concrete application, we study the twisted catenoid (graphene wormhole) acting as a geometric bottleneck. Crucially, we introduce a phenomenological structural deformation to model the twist-induced radial contraction at the constriction. By mapping the longitudinal coordinate $z$ to the meridional arc-length $s$ via an analytical perturbative approach, we obtain a 1D Klein-Gordon-like equation governed by a geometry-dependent squared effective potential. Finally, employing a rigorous numerical scattering method, we evaluate the inter-leaf transmission probability, revealing that the torsion-induced radial contraction significantly amplifies the chiral potential barrier and strongly suppresses quantum tunneling.

This paper is organized as follows. In Sec.~\ref{sec: Theoretical Framework}, we establish the theoretical framework by reviewing the curved massless Dirac equation, the tetrad formalism, and the geometric spin connection. Sec.~\ref{sec: Non-Linearly deformed objects} is dedicated to the study of non-linearly deformed geometries; we formulate a general description for arbitrary axisymmetric surfaces. In Sec.~\ref{sec:Graphene_Worm}, we apply this general geometric analysis to the twisted graphene wormhole, incorporating the twist-induced radial contraction, analytically deriving the squared effective potential, and solving the numerical scattering problem. Finally, in Sec.~\ref{sec:Conclusion}, we present our final remarks and perspectives.
\section{Theoretical Framework}
\label{sec: Theoretical Framework}

The formalism presented below applies to materials exhibiting a linear dispersion relation, such as graphene. We adopt an approach analogous to that developed in Ref.~\cite{de2007charge}. This geometric approach has been successfully employed to study relativistic fermions in fullerenes \cite{gonzalez1993electronic}. Furthermore, leveraging the mathematical analogy between the theory of topological defects in solids and the tetrad formalism of general relativity \cite{dzyaloshinskii1980poisson}, this formalism has been extensively applied to calculate the charge response to geometric defects in graphene \cite{de2007charge}.

The massless Dirac equation in a general (2+1)D curved spacetime is given by:
\begin{equation}
    i\hbar\gamma^\mu \left( \partial_\mu + \Omega_\mu \right) \psi = 0,
    \label{eq:curved_dirac}
\end{equation}
where $\Omega_\mu(x)$ is the spin connection of the spinor field \cite{collas2019dirac}. To adapt the local Lorentz transformations to the curved manifold, we employ the standard tetrad (or \textit{vielbein}) formalism $e^a_\mu(x)$, which establishes a local inertial frame at each point on the surface. The tetrads relate the general curved metric $g_{\mu\nu}$ to the flat Minkowski metric $\eta_{ab}$ via $g_{\mu\nu} = e_{\mu}^a e_{\nu}^b \eta_{ab}$. The curved-space gamma matrices $\gamma^\mu$ satisfy the generalized Clifford algebra $\{ \gamma^\mu, \gamma^\nu \} = 2g^{\mu\nu}(x)$. For condensed matter systems like graphene, these curved gamma matrices are rescaled by the Fermi velocity, such that $\gamma^\mu = \left( \gamma^0, v_F \gamma^i \right)$ with $i=1,2$ \cite{de2007charge}.

Finally, the spin connection $\Omega_\mu$, which acts as a geometry-induced pseudomagnetic gauge field ensuring the covariance of the spinor derivative, is calculated directly from the tetrads. In a torsion-free spacetime compatible with standard gravity, it is given by $\Omega_{\mu} = \frac{1}{8}\omega_\mu^{ab}\left[\gamma_a,\gamma_b\right]$, where $\omega_\mu^{ab} = e_\nu^a\left(\partial_\mu e^{b\nu}+\Gamma_{\mu\lambda}^\nu e^{b\lambda}\right)$ are the spin connection coefficients and $\Gamma^{\lambda}_{\mu \sigma}$ are the standard Christoffel symbols derived from the metric tensor \cite{nakahara2018geometry,collas2019dirac}. The central objective of this investigation is to apply this formalism to twisted surfaces, starting from the metric $g_{\mu\nu}$ derived either from the geometric definition $g_{\mu\nu}=\partial_\mu\vec{r}\cdot \partial_\nu\vec{r}$, or from the \textit{strain tensor} $u_{ij}$ \cite{soutas2012elasticity} via the relation $g_{ij}=g_{ij(0)}+2u_{ij}$.

\section{Twisted geometry and dynamics}
\label{sec: Non-Linearly deformed objects}

Before addressing the twisted graphene wormhole surface, in this section, we derive the effective eletronic dynamics for arbitrary twisted surfaces of revolution where the radius $R(z)$ varies along the longitudinal axis. This extension provides a unified description for any azimuthally symmetric geometry described in cylindrical coordinates.
Also, we assume a total twist angle given by $\phi = \alpha(z)z$, making $\alpha(z)$ the \textit{average} twist rate along the $z$ coordinate. Consequently, the function $f(z) = d\phi/dz = \alpha(z) + z\alpha'(z)$ represents the `instantaneous' local twist rate. Using the spatial parametrization $\vec{r}(z,\phi)=(R(z)\cos{(\phi+\alpha(z)z)},R(z)\sin{(\phi+\alpha(z)z)},z)$, and emphasizing that we are interested in the full $(2+1)$D spacetime, the metric using the signature $(+,-,-)$ is given by:
\begin{equation}
   \label{eq:metric_nonlin_obj_2+1}
    (g_{\mu\nu})=
        \begin{pmatrix}
            1 & 0 &  0 \\
            0 & -R(z)^2 & -R(z)^2f(z) \\
            0 & -R(z)^2f(z) & -\left(1+R(z)^2f(z)^2+R'(z)^2\right)
        \end{pmatrix}.
\end{equation}
Note that if we assume $R(z) = $const., we find the $(2+1)$D metric for the twisted cylinder case, analogous to the spatial $g_{ij}$ shown in Ref.\cite{delgado2026non}. From Eq.\eqref{eq:metric_nonlin_obj_2+1}, we determine the tetrad fields satisfying the fundamental relation $g_{\mu\nu}=e^a_\mu e^b_\nu \eta_{ab}$~\cite{de2007charge}:
\begin{equation}
   \label{eq:tetrades_nonlin_obj_2+1}
    (e_\mu^a)=
        \begin{pmatrix}
            1 & 0 &  0 \\
            0 & R(z) & 0 \\
            0 & R(z)f(z) & \sqrt{1+R'(z)^2}
        \end{pmatrix},
\end{equation}
with the corresponding inverse tetrads:
\begin{equation}
   \label{eq:inv_tetrades_nonlin_obj_2+1}
    (e^\mu_a)=
        \begin{pmatrix}
            1 & 0 &  0 \\
            0 & \frac{1}{R(z)} & 0 \\
            0 & \frac{-f(z)}{\sqrt{1+R'(z)^2}} & \frac{1}{\sqrt{1+R'(z)^2}}
        \end{pmatrix}.
\end{equation}
Calculating the spin connection coefficients $\omega_\mu^{ab}$, we find the non-vanishing components to be:
\begin{equation}
\label{eq:spin_connec_coef}
    \omega_\phi^{12}=-\frac{R'(z)}{\sqrt{1+R'(z)^2}},\quad \omega_z^{12}=-\frac{f(z)R'(z)}{\sqrt{1+R'(z)^2}}.
\end{equation}
Recalling the antisymmetry property $\omega_\mu^{ab}=-\omega_\mu^{ba}$, the spin connection spinor $\Omega_\mu(x)$ is obtained via:
\begin{equation}
\label{eq:Spin_con_12}
\Omega_\mu(x)=\frac{1}{2}\omega_\mu^{12}\gamma_1\gamma_2.
\end{equation}
As representations in $(2+1)$D are unitarily equivalent, without loss of generality, we choose a set of flat gamma matrices compatible with the Clifford algebra $\{\gamma^a, \gamma^b\} = 2\eta^{ab}$ \cite{de2007charge}. For a two-component pseudospinor, a suitable choice is:
\begin{equation}
    \gamma^0 = \sigma^z, \quad \gamma^1 = -i\sigma^x, \quad \gamma^2 = -i\sigma^y.
\end{equation}
Using this representation, we obtain the identity $\gamma_1\gamma_2 = \gamma^1\gamma^2 = (-i\sigma^x)(-i\sigma^y) = -i\sigma^z$. Therefore, we have $\Omega_{\mu}(x)=(-i/2)\omega_{\mu}^{12}\sigma^z$. Explicitly, the components are: 
\begin{equation}
    \label{eq:spin_con_general}
    \Omega_{\phi}=\frac{i}{2}\frac{R'(z)}{\sqrt{1+R'(z)^2}}\sigma^z,\quad \Omega_{z}=\frac{i}{2}\frac{f(z)R'(z)}{\sqrt{1+R'(z)^2}}\sigma^z.
\end{equation}
It is important to highlight that in the absence of torsion ($f=0$), only $\Omega_\phi$ remains present.
Defining the metric factor $S(z)=\sqrt{1+R'(z)^2}$, we derive the curved massless Dirac equation:
\begin{equation}
\label{eq:Dirac_curved_obj}
    i\hbar \partial_t \Psi=i\hbar v_F\left[\frac{\sigma^x}{S(z)}\left(\partial_z-f(z)\partial_\phi+\frac{R'(z)}{2R(z)}\right)-\frac{\sigma^y}{R(z)}\partial_\phi\right]\Psi,
\end{equation}
which yields the Hamiltonian $H$:
\begin{equation}
\label{eq:Hamiltonian_curved_obj}
    H=\hbar v_F\left[\frac{i\sigma^x}{S(z)}\left(\partial_z-f(z)\partial_\phi+\frac{R'(z)}{2R(z)}\right)-\frac{i\sigma^y}{R(z)}\partial_\phi\right].
\end{equation}
In matrix component form, this Hamiltonian is written as:
\begin{equation}
\label{eq:Hamiltonian_curved_obj_comp}
    H=\hbar v_F
    \begin{pmatrix}
        0 & \frac{i}{S}\left(\partial_z-f\partial_\phi+\frac{R'}{2R}\right)-\frac{1}{R}\partial_\phi \\
        \frac{i}{S}\left(\partial_z-f\partial_\phi+\frac{R'}{2R}\right)+\frac{1}{R}\partial_\phi & 0
    \end{pmatrix}.
\end{equation}
The effective Hamiltonian in Eq. (\ref{eq:Hamiltonian_curved_obj_comp}) shares the fundamental anti-diagonal structure and the characteristic geometry-induced connection terms (such as $R^\prime/2R$) observed in previous studies of curved Dirac materials \cite{watanabe2015electronic}. Remarkably, in the untwisted limit ($f(z)=0$), our model exactly recovers the Hamiltonian derived by Watanabe \textit{et al.} for the helicoid, up to a unitary transformation. To investigate stationary states of the axisymmetric surfaces, we apply the separation of variables $\Psi_{A,B}(x)=e^{im\phi}e^{-i\varepsilon t/\hbar}Z_{A,B}(z)$, where $m=l\pm\frac{1}{2}=\pm\frac{1}{2},\pm\frac{3}{2}, \dots$ is the total angular momentum \cite{villalba2001energy}.
The effective Hamiltonian acting on the longitudinal components $Z_{A,B}(z)$ becomes:
\begin{equation}
\label{eq:Hamiltonian_curved_obj_comp_z}
    H_{\text{eff}}=\hbar v_F
    \begin{pmatrix}
        0 & -\frac{im}{R}+\frac{mf}{S}+\frac{i}{S}\left(\partial_z+\frac{R'}{2R}\right) \\
        \frac{im}{R}+\frac{mf}{S}+\frac{i}{S}\left(\partial_z+\frac{R'}{2R}\right) & 0
    \end{pmatrix}.
\end{equation}
This leads to the coupled equations of motion:
\begin{align}
    \left[-\frac{im}{R}+\frac{mf}{S}+\frac{i}{S}\left(\partial_z+\frac{R'}{2R}\right) \right]Z_B(z)&=\epsilon Z_A(z),\\
    \left[\frac{im}{R}+\frac{mf}{S}+\frac{i}{S}\left(\partial_z+\frac{R'}{2R}\right) \right]Z_A(z)&=\epsilon Z_B(z),
\end{align}
where $\epsilon=\varepsilon/(\hbar v_F)$ is the wavenumber. To eliminate the first-derivative term proportional to $R'/2R$, we introduce the transformation $Z_{A,B}(z)=R(z)^{-1/2}\chi_{A,B}(z)$ (assuming $R(z) \neq 0$), resulting in:
\begin{align}
    \left(-\frac{im}{R}+\frac{mf}{S}+\frac{i}{S}\frac{d}{dz}\right)\chi_B(z)&=\epsilon \chi_A(z),\\
    \left(\frac{im}{R}+\frac{mf}{S}+\frac{i}{S}\frac{d}{dz}\right)\chi_A(z)&=\epsilon \chi_B(z).
\end{align}
Further simplification is achieved by removing the torsion term $mf/S$ via the gauge transformation $\chi_{A,B}(z)=\exp{\left(im\int_0^zf(\xi)d\xi\right)}u_{A,B}(z)$. This yields the reduced system:
\begin{equation}
\begin{split}
\label{eq:final_EDO_z}
    \left(\frac{m}{R}-\frac{1}{S}\frac{d}{dz}\right)u_B(z)&=i\epsilon u_A(z),\\
    \left(-\frac{m}{R}-\frac{1}{S}\frac{d}{dz}\right)u_A(z)&=i\epsilon u_B(z).
\end{split}
\end{equation}
Prior to decoupling these equations, we express the full spatial spinor reconstruction as:
\begin{equation}
\label{eq:general_solut_A,B}
    \psi_{A,B}(\phi,z)=\frac{\exp{\left[im\left(\phi+\int_0^zf(\xi)d\xi\right)\right]}}{\sqrt{R(z)}}u_{A,B}(z).
\end{equation}
Decoupling the system for $u_A(z)$ leads to the second-order differential equation:
\begin{equation}
    \label{eq:EDO_u_A(z)}
    u''_A-\frac{S'}{S}u'_A+\left[\epsilon^2S^2-\left(\frac{S}{R}\right)^2\left(m^2+\frac{R'}{S}m\right)\right]u_A=0.
\end{equation}
Finally, we perform the rescaling $u_A(z)=\sqrt{S(z)}v_A(z)$. The general solution for component A is thus:
\begin{equation}
  \label{eq:general_solut_A}
\psi_A(\phi,z)=\exp{\left[im\left(\phi+\int_0^zf(\xi)d\xi\right)\right]}\sqrt{\frac{S(z)}{R(z)}}v_A(z).
\end{equation}
Remarkably, the twist induces the appearance of a geometric phase, identical to those studied in \cite{delgado2026non}. Furthermore, $v_A(z)$ satisfies an equation similar to a Schrödinger-type equation:
\begin{equation}
    \label{eq:EDO_v_A(z)}
    v''_A(z)+\left[\frac{1}{2}\frac{S''}{S}-\frac{3}{4}\left(\frac{S'}{S}\right)^2+\epsilon^2S^2-\left(\frac{S}{R}\right)^2\left(m^2+\frac{R'}{S}m\right)\right]v_A(z)=0,
\end{equation}
with $\epsilon=\varepsilon/(\hbar v_F)$, $m=\pm\frac{1}{2},\pm\frac{3}{2},\dots$, and $S(z)=\sqrt{1+R'(z)^2}$. This formalism is applicable to any surface of revolution with radial profile $R(z)$. It is worth highlighting that $\epsilon^2S^2=\frac{\varepsilon^2}{\hbar^2\left(\frac{v_F}{\sqrt{1+R'(z)^2}}\right)^2}=\frac{\varepsilon}{\hbar \tilde{v_F}(z)}$, where the term
\begin{equation}
    \label{eq:position_Fermi_velocity}
	\tilde{v_F}(z)=\frac{v_F}{\sqrt{1+R'(z)^2}},
\end{equation}
can be understood as a \textit{position-dependent Fermi velocity} \cite{de2007charge}.
\section{Twisted Graphene Wormhole}
\label{sec:Graphene_Worm}
\subsection{Undeformed Catenoid}

Having established the general effective Hamiltonian for twisted surfaces of revolution, we now focus our analysis on the specific problem of the catenoid bridge. This minimal surface can serve as a model of geometric constrictions in nanotubes and as a bridge between two graphene leaves~\cite{silva2024strain}.
\begin{figure}[htbp]
    \centering
    \includegraphics[width=0.8\linewidth]{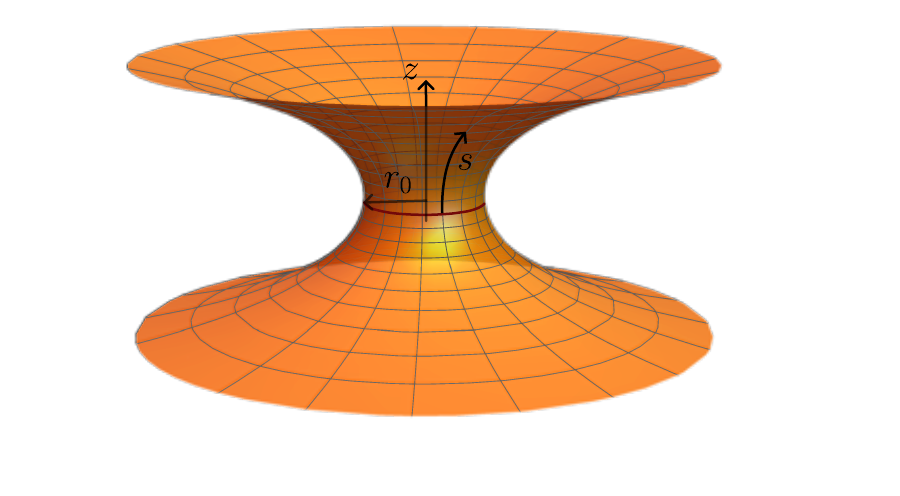}
    \caption{\textbf{Coordinate Systems on the Catenoid.} Schematic representation of the catenoid surface highlighting the relationship between the two relevant longitudinal coordinates: the standard axial coordinate $z$ (straight vertical path), and the meridional arc length coordinate $s$ (curved path along the surface). The parameter $r_0$ denotes the minimum radius at the geometric bottleneck ($z=0, s=0$). Mapping the effective 1D electron dynamics from $z$ to the path $s$ absorbs the local metric variations, allowing the system to be described by a conservative Schrödinger-like equation.}
    \label{fig:Catenoid_Coordinates}
\end{figure}

The catenoid is defined by the parametrization $\vec{r}(\phi,z)=\left(R(z)\cos\phi, R(z)\sin\phi, z\right)$, where $\phi \in [0, 2\pi)$ and $z \in (-\infty, \infty)$. The radial profile is given by:
\begin{equation}
 \label{eq:catenoid_param1}
    R(z) = r_0 \cosh\left(\frac{z}{r_0}\right),
\end{equation}
where $r_0$ corresponds to the minimum radius at the bottleneck ($z=0$). This specific geometry offers significant mathematical simplifications. The first derivative is $R'(z) = \sinh(z/r_0)$, which yields the metric factor:
\begin{equation}
 \label{eq:catenoid_param2}
    S(z) = \sqrt{1 + (R'(z))^2} = \cosh\left(\frac{z}{r_0}\right).
\end{equation}
A remarkable property of the catenoid is that the ratio between the metric factor and the radius is constant everywhere on the surface:
\begin{equation}
    \frac{S(z)}{R(z)} = \frac{1}{r_0}.
\end{equation}
This property drastically simplifies the coupling terms in the Dirac equation. Furthermore, the geometric quantities required to construct the effective potential are:
\begin{equation}
    \frac{S'}{S} = \frac{1}{r_0}\tanh\left(\frac{z}{r_0}\right), \quad \frac{S''}{S} = \frac{1}{r_0^2}, \quad \frac{R'}{S}=\tanh\left(\frac{z}{r_0}\right).
\end{equation}

To facilitate the initial analysis, one can introduce the dimensionless coordinate $\zeta = z/r_0$. Substituting these geometric parameters into the general equation of motion for the component $v_A$ derived in Eq.~(\ref{eq:EDO_v_A(z)}), we obtain:
\begin{equation}
    \label{eq:EDO_vA_Catenoid}
    \frac{d^2 v_A}{d\zeta^2} + \left[ (\epsilon r_0)^2 \cosh^2 \zeta - \left( m^2 + \frac{1}{4} + m \tanh \zeta - \frac{3}{4} \text{sech}^2 \zeta\right) \right] v_A(\zeta) = 0.
\end{equation}

While Eq.~(\ref{eq:EDO_vA_Catenoid}) provides a mathematically valid description of the dynamics, interpreting the spatially modulated kinetic term $(\epsilon r_0)^2 \cosh^2 \zeta$ can be unintuitive within the standard framework of stationary scattering problems. To circumvent this and cast the dynamics into a strictly conservative 1D scattering form, it is highly advantageous to map the system from the axial coordinate $z$ to the arc length (meridional) coordinate $s$, defined along the surface profile as illustrated in Fig.~\ref{fig:Catenoid_Coordinates}. 

The mapping between these two coordinates is given by $s = r_0\sinh(z/r_0)$ and $z = r_0\sinh^{-1}(s/r_0)$. To perform this coordinate transformation smoothly, it is more convenient to start from the unscaled Eq.~(\ref{eq:EDO_u_A(z)}), which for the catenoid reads: 
\begin{equation}
    \label{eq:EDO_u_A(z)_catenoid}
    u''_A - \frac{1}{r_0}\tanh\left(\frac{z}{r_0}\right)u'_A + \left[\epsilon^2\cosh^2\left(\frac{z}{r_0}\right) - \frac{1}{r_0^2}\left(m^2+m\tanh\left(\frac{z}{r_0}\right)\right)\right]u_A = 0.
\end{equation}
By applying the chain rule for the $s$ coordinate, the first-derivative term is exactly canceled by the metric transformation, and Eq.~(\ref{eq:EDO_u_A(z)_catenoid}) elegantly reduces to:
\begin{equation}
    \label{eq:EDO_u_A(s)_catenoid}
    \frac{d^2u_A}{ds^2} + \left[ \epsilon^2 - \left( \frac{m^2}{r_0^2+s^2} + \frac{ms}{(r_0^2+s^2)^{3/2}} \right) \right] u_A(s) = 0.
\end{equation}

Equation~(\ref{eq:EDO_u_A(s)_catenoid}) is a standard \textit{Klein-Gordon-like} equation featuring a constant energy term $\epsilon^2$ and an effective geometric potential, a result consistent with the findings in Refs.~\cite{watanabe2015electronic,silva2024strain}. It is crucial to highlight why our results for the catenoid match Watanabe's results for the helicoid. In differential geometry, the catenoid and the helicoid belong to the same associated family of minimal surfaces; they are locally isometric. By introducing continuous torsion, the catenoid can be mapped into a helicoid via Bour's deformation, strictly preserving the Gaussian curvature~\cite{caddeo2022bour}.

It is important to establish how this coordinate transformation adapts the full wavefunction structure defined in Eq.~(\ref{eq:general_solut_A,B}). Rewriting the geometric normalization and the phase integral in terms of the arc length $s$, the full pseudospinor components become:
\begin{equation}
\label{eq:solut_A,B(s)_catenoid}
    \psi_{A,B}(\phi,s) = \frac{\exp\left[ im \left( \phi + r_0 \int^s \frac{f(\xi)}{\sqrt{r_0^2+\xi^2}} d\xi \right) \right]}{\left(r_0^2+s^2\right)^{1/4}} u_{A,B}(s),
\end{equation}
where the effective torsion function in the meridional coordinate is mapped as $f(s) = \alpha(s) + \sqrt{r_0^2+s^2}\sinh^{-1}\left(\frac{s}{r_0}\right)\alpha'(s)$.

\begin{figure}[htbp]
    \centering
    \includegraphics[width=0.8\linewidth]{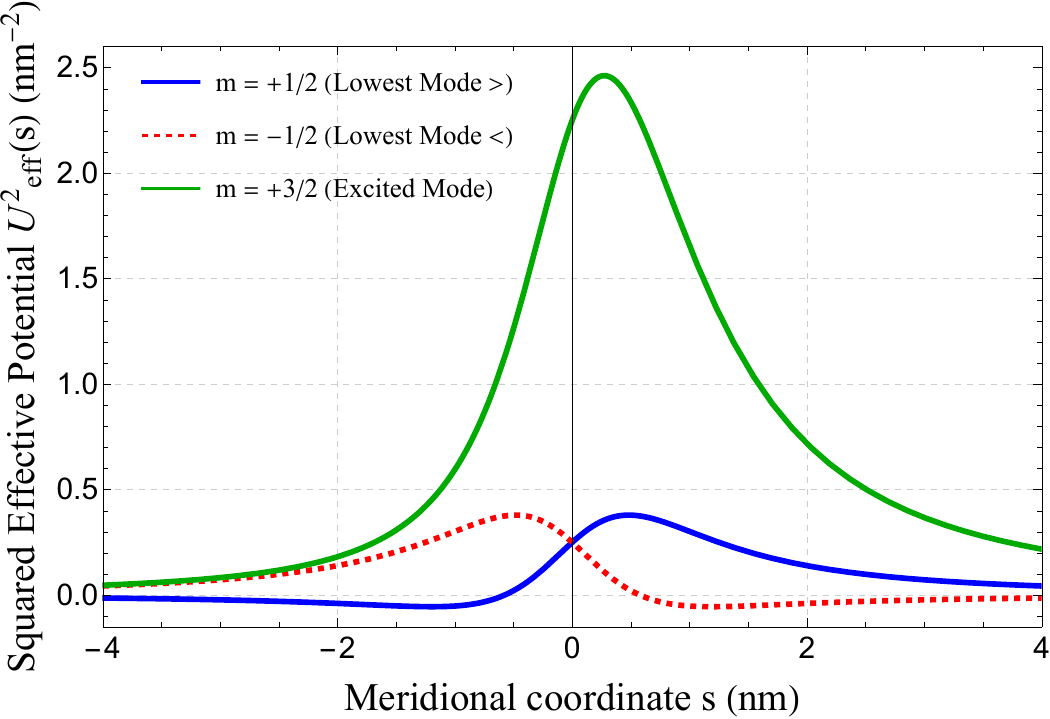} 
    \caption{\textbf{Spin-Dependent Squared Effective Potential.} The profile of the geometric potential $U_{eff}^2(s)$ is mapped along the meridional coordinate $s$ for the lowest modes $m=\pm 1/2$ and the excited mode $m=+3/2$. The geometry-induced chiral asymmetry is clearly visible: the $m=+1/2$ state (solid blue line) exhibits a higher potential barrier for $s>0$, whereas the $m=-1/2$ state (dashed red line) encounters the barrier for $s<0$. Parameters: $r_0=1$nm.}
    \label{fig:Squared_Effective_Potential}
\end{figure}

The term $\epsilon^2$, represents the squared energy parameter (or the square wavenumber) with dimensions of inverse squared length. The expression enclosed in brackets in Eq.~(\ref{eq:EDO_u_A(s)_catenoid}) dictates the effective 1D dynamics along the catenoid. From this, we can readily extract the squared effective potential $U^2_{\text{eff}}(s)$, which explicitly captures the non-trivial coupling between the fermion's pseudospin and the background geometry. Notably, this geometric potential exhibits a clear local chiral asymmetry with respect to the total angular momentum quantum number $m$:
\begin{equation}
    \label{eq:square_potential(s)_catenoid}
    U^2_{\text{eff}}(s) = \underbrace{\frac{m^2}{r_0^2+s^2}}_{\text{Centrifugal barrier}} + \underbrace{\frac{ms}{(r_0^2+s^2)^{3/2}}}_{\text{Chiral asymmetry}}.
\end{equation}
The spatial profile of this effective potential is illustrated in Fig.~\ref{fig:Squared_Effective_Potential}.

The effective potential derived in Eq.~(\ref{eq:square_potential(s)_catenoid}) exhibits distinct physical features induced by the interplay between the fermion's pseudospin and the curved geometry.
The potential term $\frac{ms}{(r_0^2+s^2)^{3/2}}$ is an odd function of the meridional coordinate $s$, explicitly breaking the spatial parity symmetry ($\mathcal{P}$) for a given spin state. However, the system obeys a combined reflection symmetry, satisfying 
\begin{equation}
    U^2_{\text{eff}}(s, m) = U^2_{\text{eff}}(-s, -m).
\end{equation}
This property implies that an electron with positive angular momentum ($m>0$) moving towards $s>0$ experiences the exact same potential landscape as an electron with negative angular momentum ($m<0$) moving towards $s<0$.

\begin{figure}[htbp]
    \centering
    \includegraphics[width=0.8\linewidth]{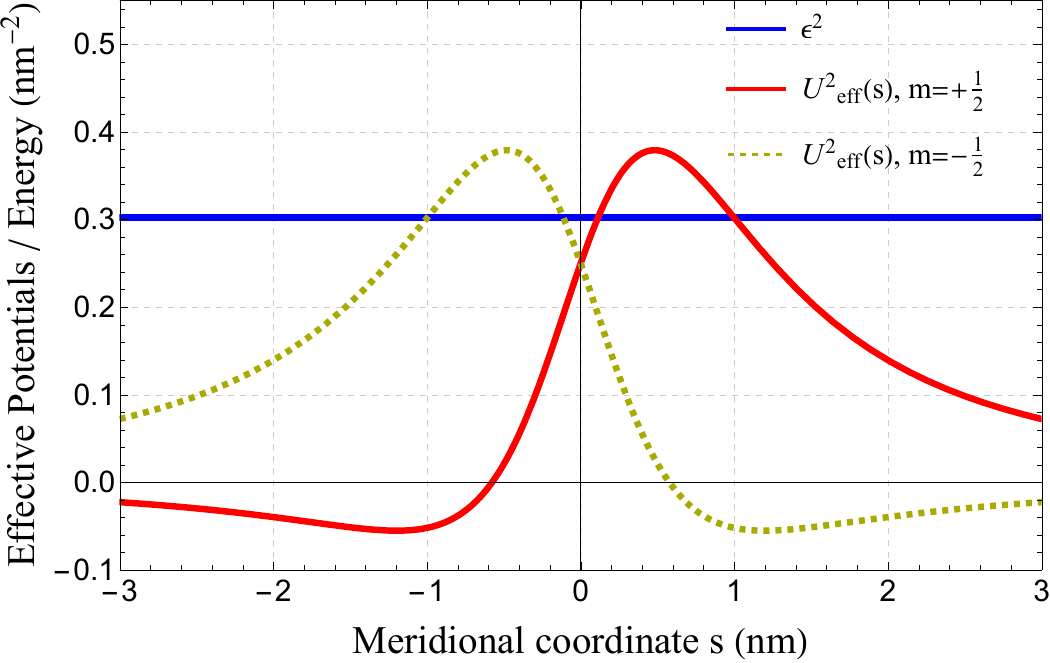} 
    \caption{\textbf{Energy Landscape and Tunneling Barrier.} Comparison between the constant squared energy parameter $\epsilon^2$ (solid blue line) and the squared effective potential $U^2_{\text{eff}}(s)$ for the lowest angular momentum modes $m=\pm 1/2$. The intersections between the energy level and the potential curves define the classical turning points, effectively delimiting the classically forbidden region (tunneling barrier). The chiral nature is evident: the Spin UP projection ($m=+1/2$, solid red line) encounters a potential barrier in the right channel ($s>0$), whereas the Spin DOWN projection ($m=-1/2$, dashed yellow line) faces an identical barrier in the left channel ($s<0$). Parameters used: $r_0 = 1$\,nm and $\epsilon^2 = 0.3025\,\text{nm}^{-2}$.}
    \label{fig:Energy_Landscape_s}
\end{figure}
Concerning the asymptotic behavior far from the constriction ($|s| \gg r_0$), the geometric potential decays and can be approximated as $U^2_{\text{eff}} \approx \frac{m(m+1)}{s^2}$. In this regime, the equation of motion reduces to:
    \begin{equation}
    \label{eq:EDO_u_A(s)_catenoid_approx}
    \frac{d^2u_A}{ds^2} + \left[ \epsilon^2 - \frac{m(m+1)}{s^2}\right] u_A(s) = 0.
    \end{equation}
Equation~(\ref{eq:EDO_u_A(s)_catenoid_approx}) can be mapped to the standard Bessel differential equation \cite{arfken2005mathematical}. This becomes more evident by introducing the dimensionless variable $x = \epsilon s$, which yields $x^2\frac{d^2u_A}{dx^2} + \left[ x^2 - m(m+1)\right] u_A(x) = 0$. By applying the substitution $u_A(x) = \sqrt{x}y(x)$, the general solution for this asymptotic region is straightforwardly given by a superposition of standard Bessel functions of the first ($J_\nu$) and second ($Y_\nu$) kinds:
    \begin{equation}
        u_A(s) \approx \sqrt{s}\left[c_1 J_{\left|m+\frac{1}{2}\right|}(\epsilon s) + c_2 Y_{\left|m+\frac{1}{2}\right|}(\epsilon s)\right].
    \end{equation}
    In the strict asymptotic limit ($|s| \to \infty$), the effective geometric potential vanishes completely ($U^2_{\text{eff}} \to 0$). Consequently, the spatial curvature no longer influences the dynamics, and the equation simplifies to $u_A''(s) = -\epsilon^2 u_A(s)$. As expected for a particle escaping an asymptotically flat geometry, the solutions recover the standard plane wave form:
    \begin{equation}
     u_A(s) \approx c_1 e^{i\epsilon s} + c_2 e^{-i\epsilon s}.
    \end{equation}

\subsection{Deformed twisted catenoid}

\begin{figure}[htbp]
    \centering
    \includegraphics[width=0.8\linewidth]{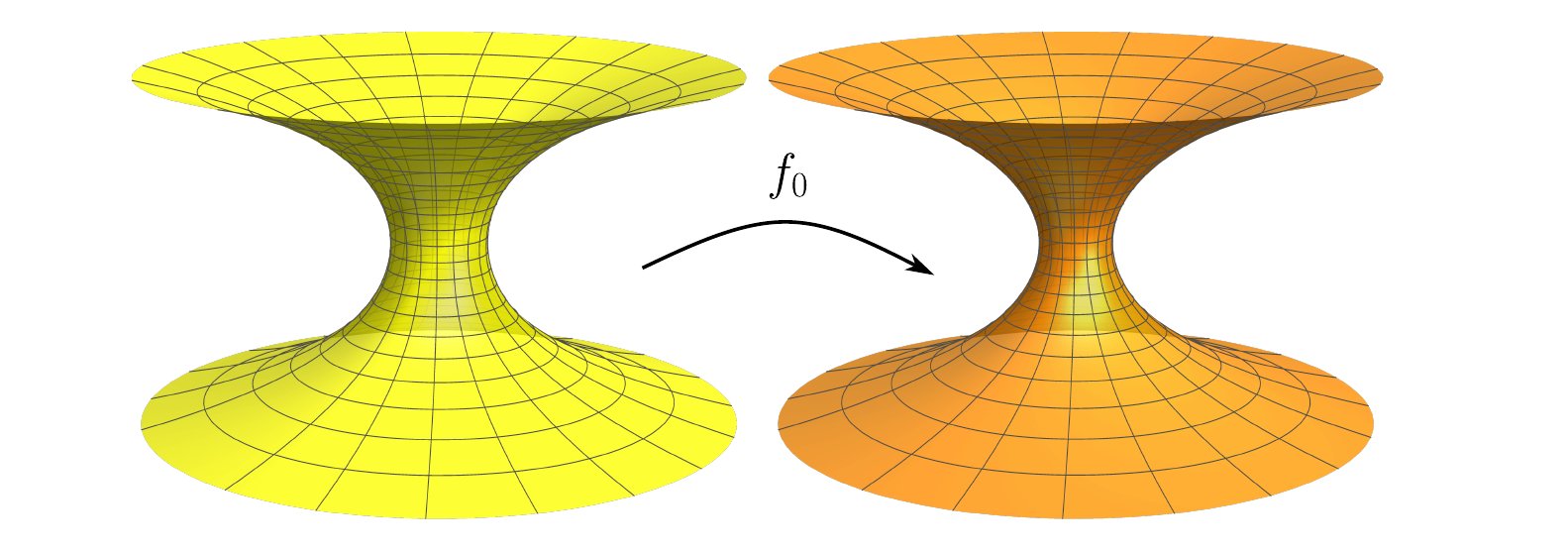}
    \caption{\textbf{Twist-induced geometric deformation of the catenoid.} The pure catenoid (left, $f_0=0$) is mapped into a modified geometry (right) under mechanical torsion. Driven by the topological ansatz, the radial profile undergoes a localized contraction explicitly concentrated around the geometric bottleneck, mimicking the elastic response of the lattice to applied twist.}
    \label{fig:deforming_catenoid}
\end{figure}

In our previous discussions, we assumed that the background geometry remains rigid under mechanical torsion. While useful, this restriction confines the influence of torsion strictly to a universal geometric phase, as demonstrated for any axisymmetric revolution surface. We now lift this constraint to provide a more realistic physical description by allowing the surface itself to deform in response to the applied twist. We start with an undeformed catenoid, which naturally constricts at its bottleneck. When subjected to torsion, the mechanical tension induces a radial contraction, meaning the surface is no longer a perfect catenoid. The deformed manifold is described by the standard axisymmetric parametrization $\vec{r}(\phi,z)=\left(R(z)\cos\phi, R(z)\sin\phi, z\right)$, with $\phi \in [0, 2\pi)$ and $z \in (-\infty, \infty)$. However, the radial profile $R(z)$ is now explicitly coupled to the \textit{uniform} twist rate $f_0$. Deriving an exact analytical form for the deformed radial profile from first principles in continuum elasticity theory is highly non-trivial and lies beyond the scope of this work. To circumvent this issue, we introduce the phenomenological \textit{geometric ansatz}:
\begin{equation}
\label{eq:deform_radial_prof}
    R(z) = r_0\left[1-(f_0 r_0)^2\text{sech}^2\left(\frac{z}{r_0}\right)\right]\cosh\left(\frac{z}{r_0}\right),
\end{equation}
where $r_0$ is the minimum radius of the unperturbed catenoid at the throat ($z=0$). This specific ansatz was chosen to satisfy fundamental physical constraints: it is dimensionally consistent, and its quadratic dependence on torsion $(f_0 r_0)^2$ preserves the structural invariance under parity inversion of the twist direction, $R(z, f_0) = R(z, -f_0)$. It is important to emphasize that at the maximum constriction ($z=0$), the radial profile reduces to $R(0)=r_0\left[1-(r_0f_0)^2\right]$. This physical boundary condition naturally restricts our analysis to a regime of moderate twist rates satisfying $|f_0|\leq 1/r_0$. Furthermore, because the $\text{sech}^2(z/r_0)$ function decays rapidly at infinity, the structural contraction acts as a localized perturbation strictly concentrated around the geometric constriction, smoothly recovering the unperturbed catenoid profile in the asymptotic regions. This geometrically localized necking effect is illustrated in Fig. \ref{fig:deforming_catenoid}.

\begin{figure}[htbp]
    \centering
    \includegraphics[width=0.8\linewidth]{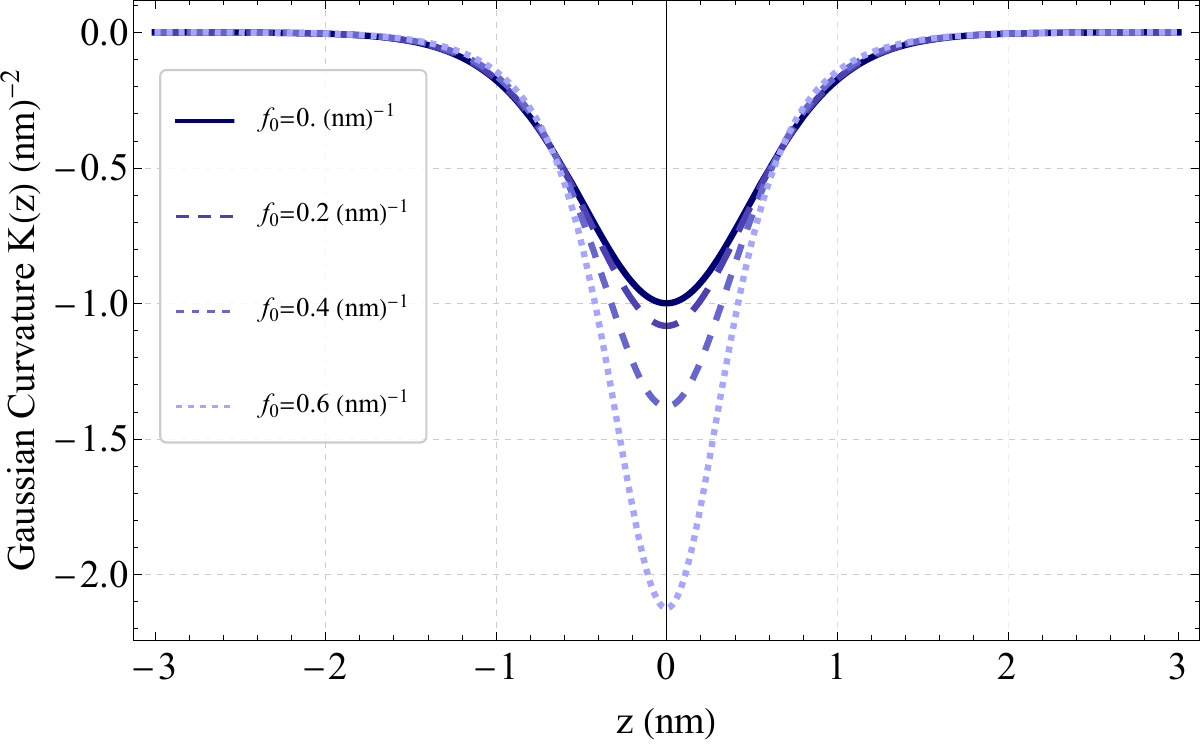} 
    \caption{\textbf{Twist-induced modification of the Gaussian curvature.} Spatial profile of the Gaussian curvature $K(z)$ along the axial coordinate $z$ for selected twist rates $f_0$. The localized radial contraction at the bottleneck ($z=0$) strongly intensifies the negative Gaussian curvature, reflecting the mechanical tightening of the graphene wormhole throat. Parameters: $r_0=1$nm.}
    \label{fig:gaussian_curvature}
\end{figure}

\begin{figure}[htbp]
    \centering
    \includegraphics[width=0.8\linewidth]{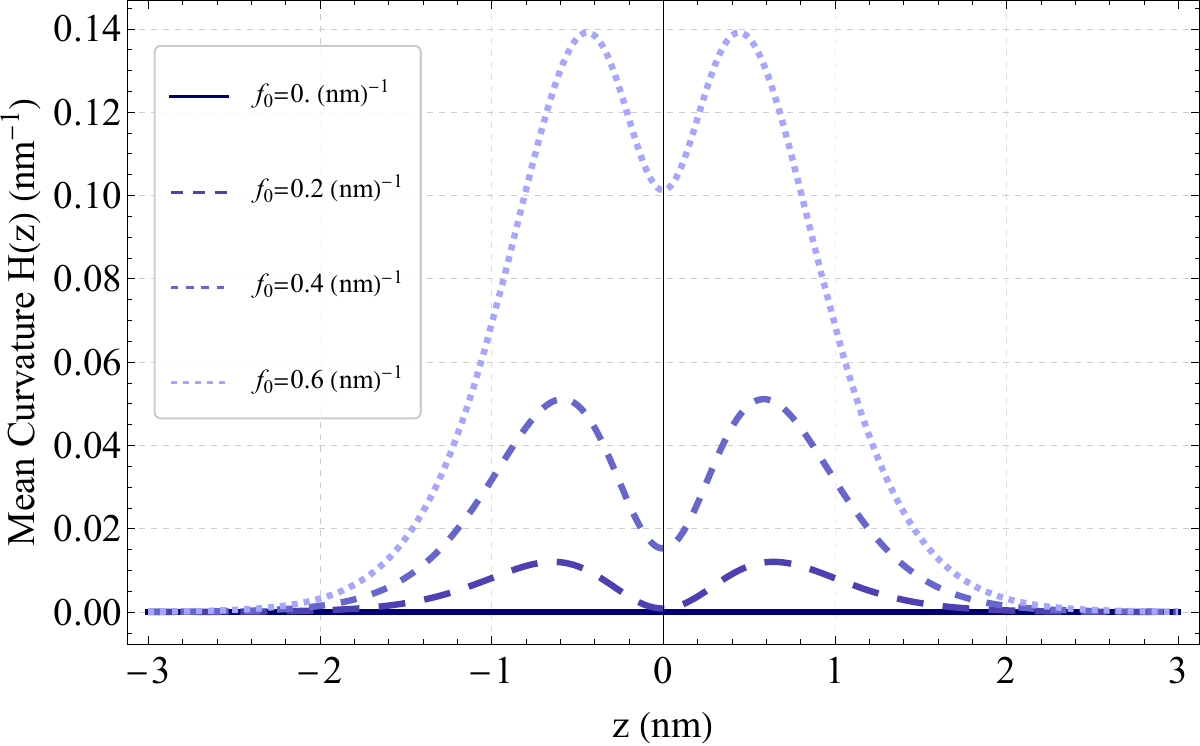} 
    \caption{\textbf{Symmetry breaking of the minimal surface.} Mean curvature $H(z)$ profile of the deformed catenoid. For the unperturbed geometry ($f_0 = 0$), the surface is perfectly minimal ($H=0$). The introduction of mechanical torsion forces the geometry to deviate from the minimal state, creating a highly localized double-peak structure around the constriction. This non-zero mean curvature is the exact geometric manifestation of the twist-induced structural deformation. Parameters: $r_0=1$nm.}
    \label{fig:mean_curvature}
\end{figure}

For an arbitrary surface, the Gaussian curvature $K$ and the mean curvature $H$ are fundamentally defined in terms of the \textit{first fundamental form} $g_{ij}$ and the \textit{second fundamental form} $h_{ij}$ as ~\cite{da1981quantum}
\begin{equation}
    K = \frac{\det(h_{ij})}{\det(g_{ij})}, \quad \text{and} \quad H = \frac{1}{2\det(g_{ij})}(g_{11}h_{22} + g_{22}h_{11} - 2g_{12}h_{12}).
\end{equation}
Using the deformed ansatz from Eq.(~\ref{eq:deform_radial_prof}) in these geometric relations, we obtain the localized curvature profiles.
To rigorously quantify the geometric impact of the twist-induced radial contraction, we evaluate the fundamental curvatures of the deformed surface. As depicted in Fig.~\ref{fig:gaussian_curvature}, the mechanical torsion significantly intensifies the negative Gaussian curvature at the bottleneck ($z=0$), reflecting the physical tightening of the wormhole throat. Furthermore, the structural deformation fundamentally alters the minimal nature of the surface. Figure~\ref{fig:mean_curvature} illustrates that the mean curvature $H(z)$ deviates from zero, developing a highly localized double-peak structure around the constriction to accommodate the mechanical stress. Crucially, in the limit of vanishing torsion ($f_0 \to 0$), the mean curvature vanishes completely ($H=0$) and the Gaussian curvature strictly reduces to the standard form $K(z) = -r_0^{-2}\text{sech}^4(z/r_0)$, flawlessly recovering the classical unperturbed minimal catenoid geometry.

Following the established framework, we investigate the effective 1D longitudinal dynamics of the pseudospinor component $u_A(z)$ via Eq. (\ref{eq:EDO_u_A(z)}). The modified radial profile naturally alters the local metric factor $S(z) = \sqrt{1+[R'(z)]^2}$, which takes the exact analytical form:
\begin{equation}
\begin{split}
    S(z) = \cosh\zeta \Big[ 1 + &2(f_0 r_0)^2 \tanh^2\zeta \text{sech}^2\zeta \\
    & + (f_0 r_0)^4 \tanh^2\zeta \text{sech}^4\zeta \Big]^{1/2}
\end{split}
\label{eq:metric_factor_deformed}
\end{equation}
where $\zeta = z/r_0$. Substituting these deformed geometric quantities back into Eq.(\ref{eq:EDO_u_A(z)}) yields a second-order differential equation of the form $u_A''(z) + p(z)u_A'(z) + q(z)u_A(z) = 0$. However, the exact expressions for the coefficients $p(z)$ and $q(z)$ become algebraically cumbersome and obscure the underlying scattering physics. To restore a strictly conservative Klein-Gordon-like form, we map the longitudinal coordinate $z$ to the meridional arc-length coordinate $s$ using the relation $ds = S(z)dz$. This coordinate transformation naturally absorbs the first-derivative term, simplifying the effective dynamics to:
\begin{equation}
    u_A''(s) + \left[\epsilon^2 - U_{\text{eff}}^2(s)\right]u_A(s) = 0,
\end{equation}
where the exact squared effective potential governing the constrained fermions is given by:
\begin{equation}
    \label{eq:squa_eff_tors_potential}
    U_{\text{eff}}^2(s) = \frac{m^2}{R(s)^2} + \frac{R'(s)}{R(s)^2}m.
\end{equation}
Due to the algebraic complexity of the deformed metric factor $S(z)$, the exact mapping $s = \int S(z)dz$ does not admit a closed-form analytical inverse in terms of elementary functions. Furthermore, as expected, Eq.(\ref{eq:squa_eff_tors_potential}) gracefully reduces to the unperturbed purely geometric potential of Eq.(\ref{eq:square_potential(s)_catenoid}) in the limit of vanishing torsion ($f_0 \to 0$).
\begin{figure*}[htbp]
    \centering
    \begin{minipage}[b]{0.5\textwidth}
        \centering
        \includegraphics[width=\linewidth]{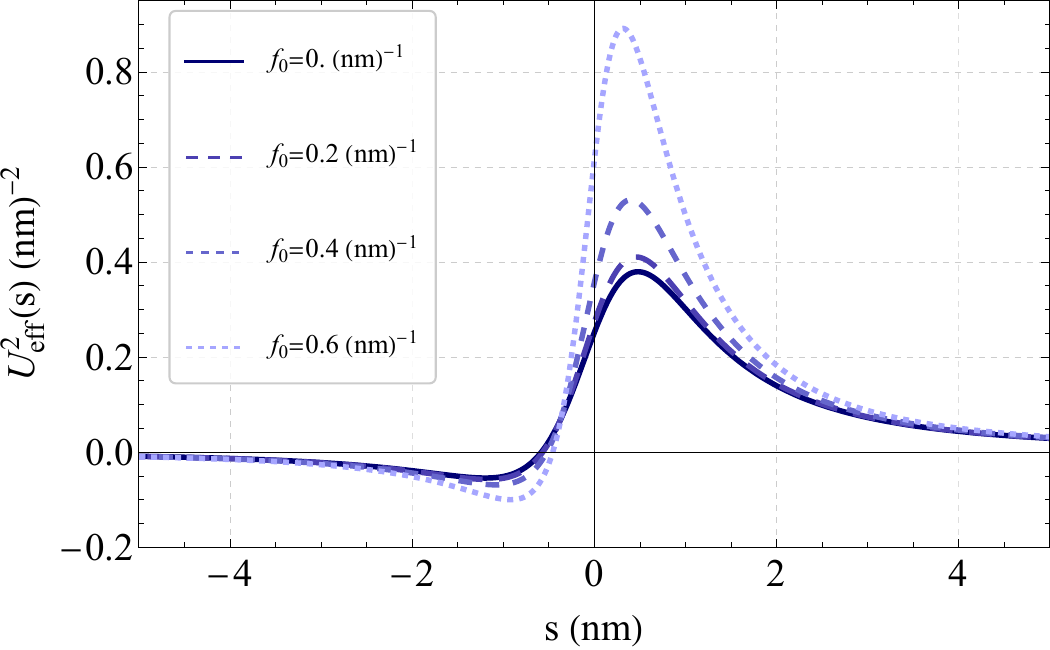}
        \vspace{0.1cm}
        \centerline{(a)}
        \label{fig:potent_var_tor}
    \end{minipage}\hfill
    \begin{minipage}[b]{0.5\textwidth}
        \centering
        \includegraphics[width=\linewidth]{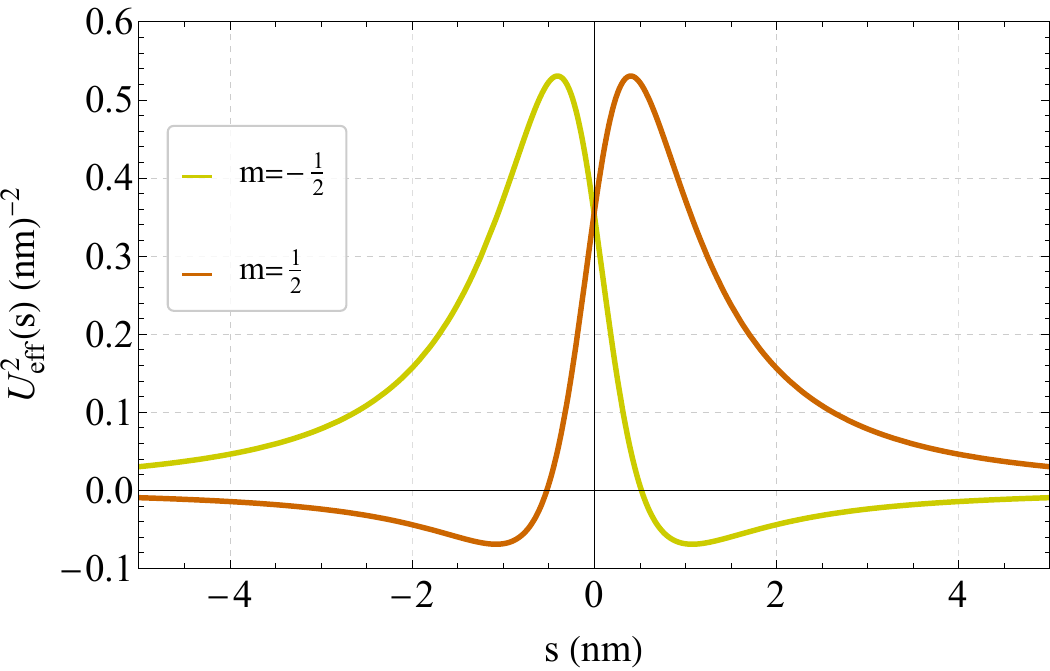}
        \vspace{0.1cm}
        \centerline{(b)}
        \label{fig:potent_assym}
    \end{minipage}
    \caption{\textbf{Effective potential of the twisted catenoid.} (a) The squared effective potential $U_{\text{eff}}^2(s)$ as a function of the meridional coordinate $s$ for the positive valley ($m = +1/2$). As the uniform twist rate $f_0$ increases, the radial contraction at the bottleneck significantly amplifies the effective potential barrier. (b) Squared effective potential landscape for fermions in opposite valleys ($m = \pm 1/2$) under a fixed torsion rate of $f_0 = 0.4 \text{ nm}^{-1}$. The spatial mirroring of the effective barriers visually confirms the preservation of the reflection symmetry $U_{\text{eff}}^2(s, m) = U_{\text{eff}}^2(-s, -m)$.}
    \label{fig:effective_potential}
\end{figure*}

This squared effective potential is plotted numerically in Fig. \ref{fig:effective_potential}. As shown in Fig. \ref{fig:potent_var_tor}, increasing the twist rate elevates the potential barrier near the constriction, explicitly demonstrating that inter-leaf transmission across the catenoid bridge is suppressed by the mechanical torsion. Furthermore, in the limit of vanishing torsion ($f_0 = 0$), we perfectly recover the unperturbed geometric potential previously depicted in Fig. \ref{fig:Energy_Landscape_s}. The local spin asymmetry is strictly preserved, as illustrated in Fig. \ref{fig:potent_assym}, proving that the geometry-induced valley polarization remains a robust topological feature even after the non-linear surface deformation.

\begin{figure}[htbp]
    \centering
    \includegraphics[width=0.8\linewidth]{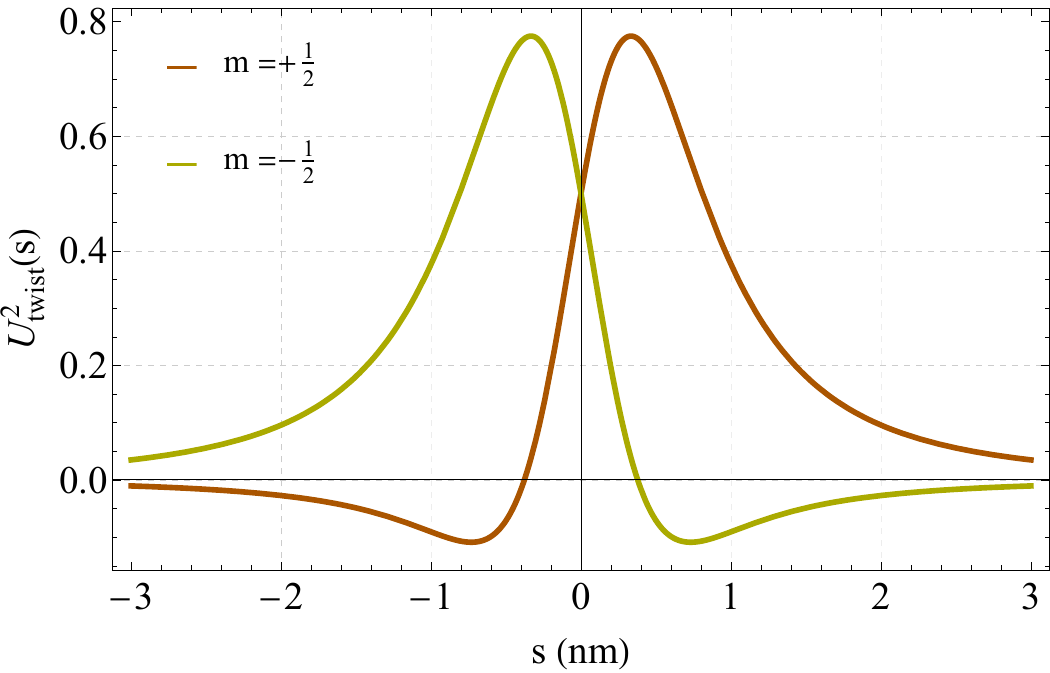} 
    \caption{\textbf{Twist-induced perturbation potential.} Spatial profile of the second-order geometric correction $U_{\text{twist}}^2(s)$ along the meridional coordinate $s$, isolated from the background geometry. The curves correspond to the opposite pseudospin valleys $m=+1/2$ (solid orange) and $m=-1/2$ (solid yellow). The perturbation rigorously preserves the chiral inversion symmetry $U_{\text{twist}}^2(s, m) = U_{\text{twist}}^2(-s, -m)$ of the unperturbed catenoid. Because both the background geometry and the twist perturbation share the exact same parity-breaking signature, the twist-induced radial contraction constructively amplifies the effective potential barrier at the bottleneck.}
    \label{fig:u2twist}
\end{figure}

As previously discussed, the exact integral mapping $s(z) = \int_0^z S(z', f_0) dz'$ does not admit a closed-form analytical inverse, which prevents us from expressing the potential in Eq.(\ref{eq:squa_eff_tors_potential}) exactly in terms of the meridional coordinate $s$. To circumvent this mathematical obstacle and extract the underlying physics, we implement a perturbative analytical approach. We begin by expanding the inverse mapping $z(s)$ with respect to the torsion parameter using the following \textit{ansatz}:
\begin{equation}
    z(s) \approx z_0(s) + f_0^2 z_2(s), \quad \text{with} \quad z_0(s) = r_0\sinh^{-1}\left(\frac{s}{r_0}\right).
\end{equation}
Our central challenge is to analytically determine the structural correction function $z_2(s)$. The starting point is to expand the deformed metric factor $S(z, f_0)$ up to second order in the twist parameter. Due to the parity invariance of our radial profile, odd powers of $f_0$ naturally vanish, yielding $S(z, f_0) \approx S_0(z) + f_0^2 S_2(z)$, where $S_0(z) = \cosh(z/r_0)$. By definition, the identity mapping $s(z(s)) = s$ must hold:
\begin{equation*}
    s = \int_0^{z(s)} S(z', f_0) dz'.
\end{equation*}
Substituting our perturbative expansions into the integral boundaries and the integrand, we obtain:
\begin{equation*}
    s \approx \int_0^{z_0(s) + f_0^2 z_2(s)} \left[ S_0(z') + f_0^2 S_2(z') \right] dz'.
\end{equation*}
By evaluating this expression via the Fundamental Theorem of Calculus and strictly retaining terms up to $O(f_0^2)$, we arrive at the consistency condition:
\begin{equation*}
    0 = f_0^2 \left[ S_0(z_0(s))z_2(s) + \int_0^{z_0(s)} S_2(z') dz' \right],
\end{equation*}
which allows us to algebraically isolate the correction term as:
\begin{equation}
    z_2(s) = - \frac{\int_0^{z_0(s)} S_2(z') dz'}{S_0(z_0(s))}.
\end{equation}
Evaluating this perturbative inversion analytically, we obtain the exact closed-form expression for $z_2(s)$:
\begin{equation}
    z_2(s) = \frac{r_0^4 \left[r_0 s - 2 \left(r_0^2+s^2\right) \tan^{-1}\left(\tanh\left(\frac{1}{2} \sinh^{-1}\left(\frac{s}{r_0}\right)\right)\right)\right]}{2 \left(r_0^2+s^2\right)^{3/2}}.
\end{equation}
With the coordinate mapping rigorously established, we can now express the squared effective potential up to second-order in the twist rate:
\begin{equation}
    U^2_{\text{eff}}(s) \approx U^2_{\text{geo}}(s) + f_0^2 U^2_{\text{twist}}(s),
\end{equation}
where the zero-order term $U^2_{\text{geo}}(s)$ perfectly matches the known unperturbed background contribution from Eq. (\ref{eq:square_potential(s)_catenoid}), explicitly given by:
\begin{equation}
    U^2_{\text{geo}}(s) = \frac{m^2}{r_0^2+s^2} + \frac{ms}{(r_0^2+s^2)^{3/2}}.
\end{equation}
The perturbation term $U^2_{\text{twist}}(s)$, when written explicitly in terms of $s$, becomes algebraically cumbersome. The plot of the perturbation correction function $U^2_{\text{twist}}(s)$ is shown in Fig.\ref{fig:u2twist}. To unveil its elegant underlying structure, we introduce the dimensionless auxiliary variable $y(s) = \sinh^{-1}(s/r_0)$. This change of variables implies $s = r_0\sinh y$ and $\sqrt{r_0^2+s^2} = r_0\cosh y$. Consequently, the pure twist correction reduces to a remarkably compact analytical form:
\begin{align}
    U^2_{\text{twist}}(y) &= \frac{1}{2}m\,\text{sech}^3y \Big[\text{gd}(y) \left(2m\tanh y - 3\text{sech}^2y + 2\right)\notag\\
    &\quad + \text{sech}^3y (2m + 5\tanh y) + 2\text{sech}y (m + \tanh y)\Big],
\end{align}
where $\text{gd}(y)$ is the well-known \textit{Gudermannian function}, defined as:
\begin{equation*}
    \text{gd}(y) = \int_0^y \frac{dt}{\cosh t}.
\end{equation*}
\subsubsection{Quantum Scattering and Numerical Analysis}

Having established that the twist-induced radial contraction significantly modifies the effective potential, a natural subsequent step is to quantify the influence of this geometric deformation on the scattering dynamics within the well-behaved moderate torsion regime. As depicted in Fig.~\ref{fig:effective_potential}, the system does not support bound states for $\epsilon^2 > 0$, thereby restricting our focus to scattering phenomena. In the asymptotic limit ($|s|\rightarrow\infty$), the effective potential vanishes, and the wave equation gracefully reduces to the free-particle form:
\begin{equation}
 \label{eq:pure_geom_assynt_eq}
    u_A''(s) = -\epsilon^2 u_A(s),
\end{equation}
where $\epsilon$ is the energy wavenumber. Consider an electron incident from the far left (or bottom) of the twisted graphene wormhole. As this massless Dirac fermion approaches the wormhole throat, it experiences an effective potential barrier strictly localized at the geometric constriction. Consequently, the asymptotic scattering states on either side of the barrier can be expressed as:
\begin{align}
    u_{L}(s) &= e^{i\epsilon s} + R e^{-i\epsilon s},\\
    u_{R}(s) &= T e^{i\epsilon s}.
\end{align} 
In this standard representation, the transmission and reflection probabilities are defined as $\mathcal{T}=|T|^2$ and $\mathcal{R}=|R|^2$, respectively. However, implementing this directly in numerical boundary value solvers is challenging since the transmitted amplitude $T$ is not known \textit{a priori}. To circumvent this mathematical obstacle, we reformulate the boundary conditions by imposing a predefined purely transmitted plane wave at the right boundary:
\begin{align}
    \tilde{u_{L}}(s) &= A e^{i\epsilon s} + B e^{-i\epsilon s},\\
    \tilde{u_{R}}(s) &= e^{i\epsilon s}.
\end{align} 
Under this formulation, the modified incident and reflected amplitudes ($A$ and $B$) yield the physical scattering probabilities $\mathcal{T}=1/|A|^2$ and $\mathcal{R}=|B|^2/|A|^2$. Numerically, we perform a reverse integration (a shooting method) starting from a sufficiently large asymptotic distance $s_{max}=100$ nm, integrating backwards to $-s_{max}$. By evaluating the numerical solution and its derivative at the left boundary, $u_{num} \equiv \tilde{u_A}(-s_{max})$ and $u_{num}' \equiv \tilde{u_A}'(-s_{max})$, we construct the linear system:
\begin{equation}
    u_{num} = A e^{-i\epsilon s_{max}} + B e^{+i\epsilon s_{max}}, \quad u_{num}' = i\epsilon A e^{-i\epsilon s_{max}} - i\epsilon B e^{+i\epsilon s_{max}}.
\end{equation}
Inverting this system provides the complex amplitudes:
\begin{equation}
    A = \frac{1}{2}e^{i\epsilon s_{max}}\left(u_{num} + \frac{u'_{num}}{i\epsilon}\right), \quad B = \frac{1}{2}e^{-i\epsilon s_{max}}\left(u_{num} - \frac{u'_{num}}{i\epsilon}\right).
\end{equation}
Knowing the explicit values of $(u_{num}, u'_{num})$, we accurately determine $(A,B)$ and, consequently, the scattering probabilities. To computationally solve the second-order differential equation, we decompose it into a coupled system of first-order equations. By defining $y_1(s)=\tilde{u_A}(s)$ and $y_2(s)=\tilde{u_A}'(s)$, we obtain the relations:
\begin{equation}
    y_2(s) = y_1'(s), \quad y_2'(s) = \left[U^2_{\text{eff}}(s) - \epsilon^2\right]y_1(s).
\end{equation}
With this set of equations and the established boundary values, the exact numerical calculation of the transmission and reflection probabilities becomes straightforward.

\begin{figure}[htbp]
    \centering
    \includegraphics[width=0.8\linewidth]{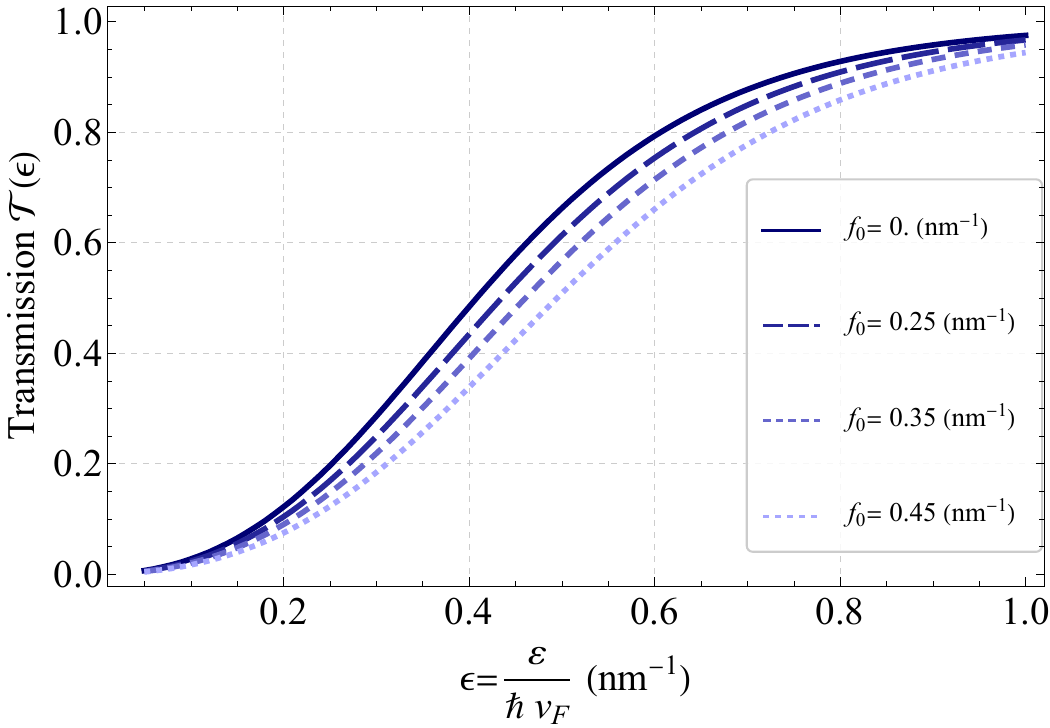} 
    \caption{\textbf{Twist-induced suppression of quantum tunneling.} Exact transmission probability $\mathcal{T}$ as a function of the incident energy wavenumber $\epsilon$ for massless Dirac fermions traversing the graphene wormhole. The curves correspond to selected uniform torsion rates $f_0 \in \{0.0, 0.25, 0.35, 0.45\} \text{ nm}^{-1}$. In the strictly geometric framework (absence of elastic lattice strain), increasing the mechanical twist enhances the local radial contraction at the throat, which directly amplifies the squared effective potential barrier and suppresses the inter-leaf transmission. Fixed structural parameters: bottleneck radius $r_0 = 1.0 \text{ nm}$ and pseudospin channel $m = +1/2$. The numerical scattering was performed by evaluating the exact analytical perturbative geometric correction up to $\mathcal{O}(f_0^2)$, ensuring the validity of the continuum limit.}
    \label{fig:pure_geom_trans}
\end{figure}

As demonstrated in Fig.~\ref{fig:pure_geom_trans}, increasing the geometric twist directly \textit{suppresses the probability} of the electron traversing the constriction. This reduced transmission is physically consistent with our previous analytical observations: within the moderate torsion regime, the radial contraction induced by $f_0$ effectively amplifies the geometric potential barrier at the throat (as previously shown in Fig.~\ref{fig:effective_potential}), thereby hindering electronic tunneling. Notably, in the absence of torsion $(f_0=0)$, our exact numerical results are in perfect quantitative agreement with those reported by Watanabe \textit{et al.}~\cite{watanabe2015electronic}, fully recovering both the functional form and the exact numerical values of the transmission probability for the unperturbed catenoid.
\begin{figure}[htbp]
    \centering
    \begin{minipage}{0.49\textwidth}
        \centering
        \includegraphics[width=\linewidth]{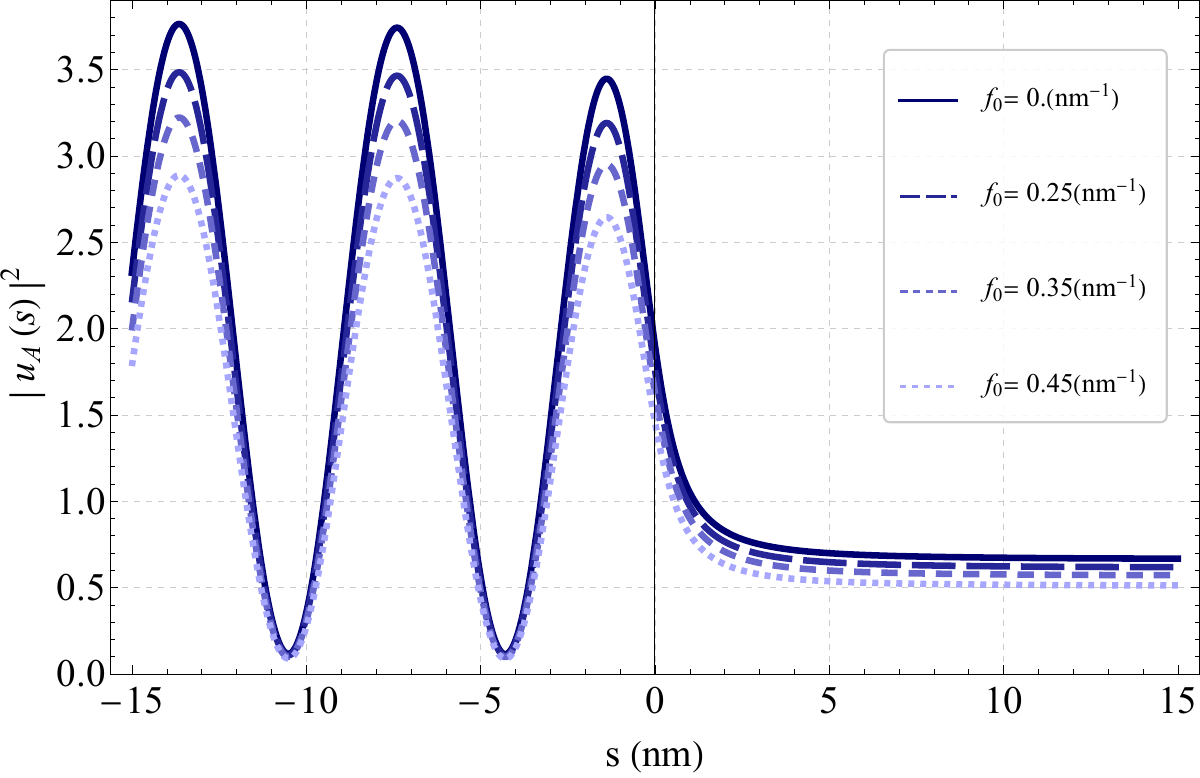}
        (a) $m = +1/2$
    \end{minipage}\hfill
    \begin{minipage}{0.49\textwidth}
        \centering
        \includegraphics[width=\linewidth]{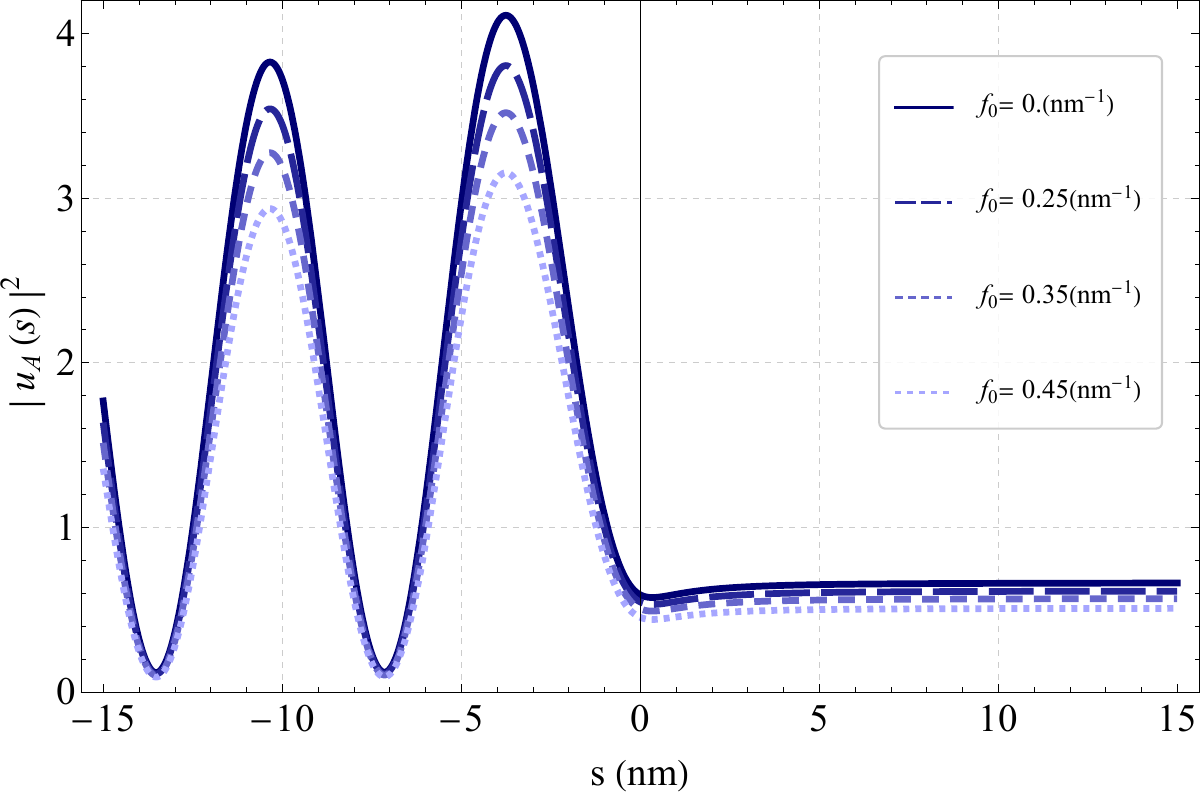}
        (b) $m = -1/2$
    \end{minipage}
    \caption{Probability density $|\tilde{u_A}(s)|^2/|A|^2$ evaluated along the meridional coordinate $s$ for a fixed incident energy parameter ($\epsilon = 0.50 \text{ nm}^{-1}$) for opposite pseudospin valleys. Panel (a) shows the dynamics for $m = +1/2$, while panel (b) illustrates the $m = -1/2$ channel. The spatial distinction between the interference patterns and barrier penetration visually highlights the geometry-induced chiral asymmetry, while the progressive lowering of the asymptotic transmission plateaus demonstrates the torsion-induced suppression of quantum tunneling in both channels.}
    \label{fig:density}
\end{figure}

\begin{figure}[htpb]
    \centering
    \begin{minipage}{0.49\textwidth}
        \centering
        \includegraphics[width=\linewidth]{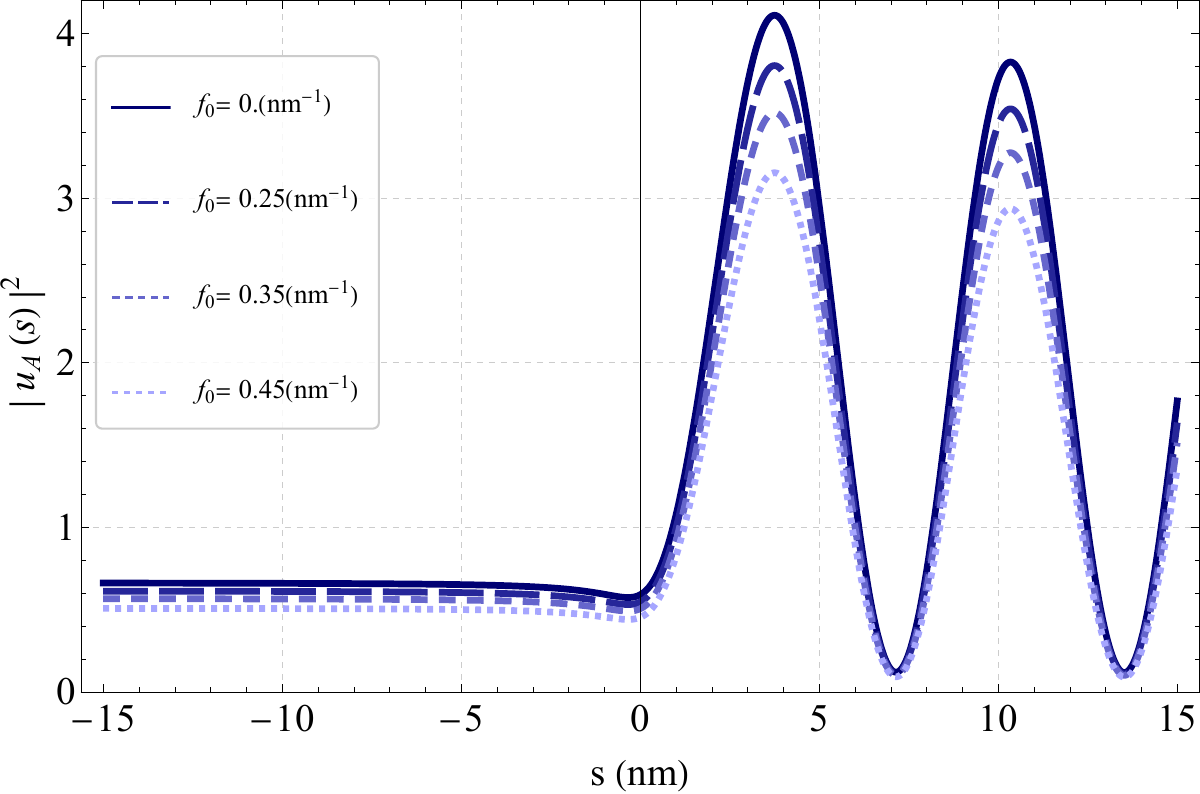}\\
        (a) $m = +1/2$
    \end{minipage}\hfill
    \begin{minipage}{0.49\textwidth}
        \centering
        \includegraphics[width=\linewidth]{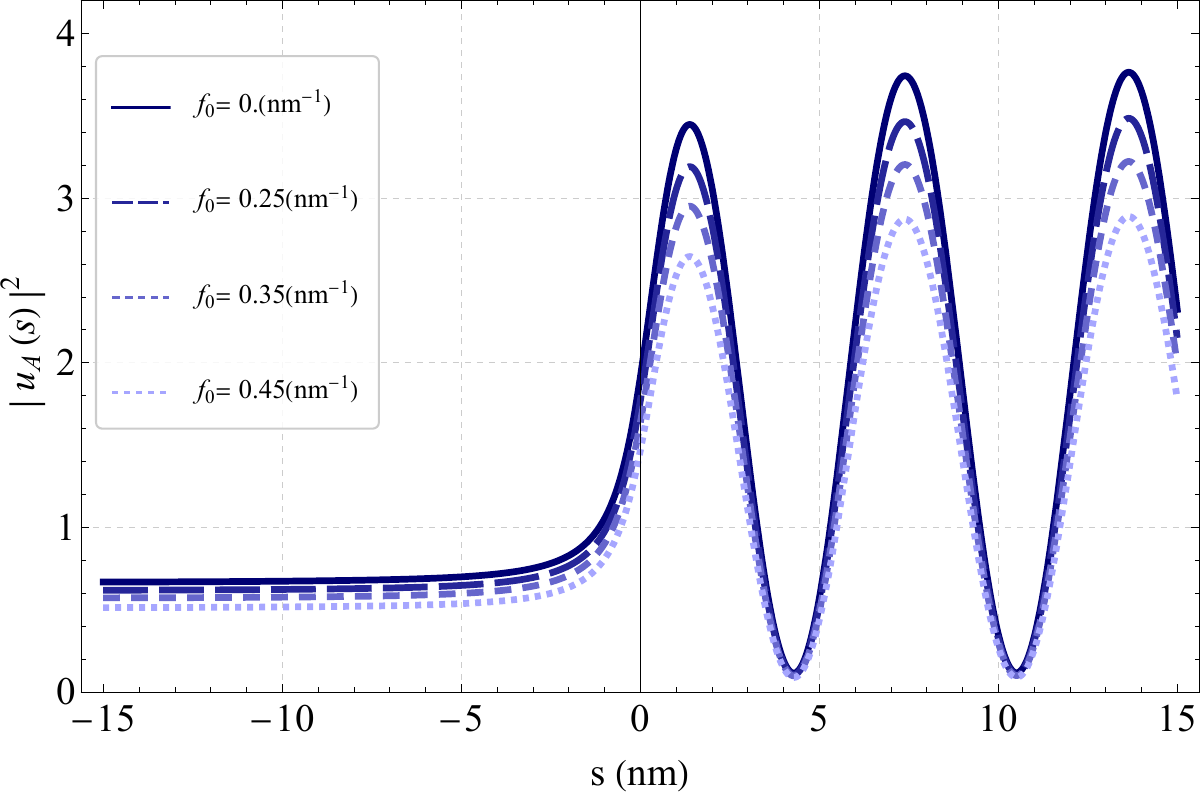}\\
        (b) $m = -1/2$
    \end{minipage}
    \caption{Probability density $|\tilde{u_A}(s)|^2/|A|^2$ evaluated along the meridional coordinate $s$ for a fixed incident energy parameter ($\epsilon = 0.50 \text{ nm}^{-1}$) for opposite pseudospin valleys. Panel (a) shows the wave incident from the right with $m = +1/2$, while panel (b) shows the wave incident from the right with $m = -1/2$, demonstrating the reversed asymptotic scattering behavior and the spatial symmetry of the effective potential.}
    \label{fig:probability_density}
\end{figure}

To complement our discussion, we illustrate the scattering behavior of a massless Dirac fermion incident from the \textit{right} in Fig. \ref{fig:probability_density}. This profile is rigorously obtained by modifying the asymptotic boundary conditions of the numerical integration, imposing a purely transmitted wave at the left extremity.

\section{Final Remarks and Perspectives}
\label{sec:Conclusion}

In this paper, we investigated the quantum dynamics of massless Dirac fermions constrained to twisted surfaces of revolution, with a special focus on the graphene wormhole. To comprehensively describe the system's dynamics, we employed the purely curved Dirac equation. While this framework provides a rigorous geometric description, it inherently neglects the microscopic changes that mechanical deformations induce in the crystal lattice. Within this purely geometric description, a ubiquitous feature across the discussed geometries is the emergence of a topological phase, presenting a clear analogy with the Aharonov-Bohm effect \cite{aharonov1959significance}. This phase depends explicitly on the torsion parameter $f(z)$ and arises from the fundamental coupling between the curved-space Dirac equation and the background geometry. Our main findings can be summarized as follows:

    First, we derived a general equation governing the effective longitudinal dynamics of a Dirac fermion on an arbitrary surface of revolution $R(z)$. We demonstrated that the pseudospinor naturally acquires a geometric phase dictated by both the surface's radial profile and its torsion, establishing this as a universal topological feature for such constrained relativistic geometries.

    Second, regarding the catenoid, we applied our general 1D formalism. To circumvent the interpretational subtleties associated with the cylindrical axial coordinate ($z$), we mapped the effective 1D equation to the meridional arc-length coordinate ($s$). This transformation revealed a squared effective potential exhibiting geometry-induced chiral symmetry breaking, a phenomenon also reported in Refs.~\cite{watanabe2015electronic,silva2024strain}. Furthermore, by analyzing the asymptotic limits ($|s|\gg r$), we found that the spatial solutions $u_A(s)$ naturally converge to a superposition of Bessel functions. 

     Finally, we investigated the physical scenario in which the twist mechanism actively contracts the surface radii around the maximum constriction. To model this, we introduced a phenomenological geometric \textit{ansatz} for the radial profile $R(z, f_0)$, directly coupling it to the twist rate. By rigorously expanding the metric and mapping the system to the meridional arc coordinate $s$, we obtained an effective 1D Klein-Gordon-like equation. This equation features a squared effective potential barrier localized at the bottleneck, which is strongly modulated by the twist parameter $f_0$. Finally, by evaluating the exact numerical scattering, we demonstrated that increasing the torsion suppresses the transmission probability across the graphene wormhole bridge, while perfectly recovering the unperturbed results of Watanabe \textit{et al.}~\cite{watanabe2015electronic} in the limit $f_0=0$.

\subsubsection*{Perspectives}

This work paves the way for several natural extensions in the study of quantum materials on curved backgrounds. First, the inclusion of external uniform magnetic fields coupled with the structural torsion would allow for a detailed investigation of Landau quantization and the possible emergence of magnetically induced localized bound states near the geometric constriction. Furthermore, exploring other types of mechanical deformations beyond uniform torsion could reveal novel geometric confinement effects. Another compelling avenue is adopting the effective tight-binding strain Hamiltonian formalism \cite{de2012space,de2013gauge}, which incorporates additional gauge fields generated by the microscopic lattice distortions—a crucial step to fully capture the electromechanical interplay in strained 2D materials. 

From a many-body perspective, extending this single-particle framework to a quantum gas of $N$ non-interacting fermions would enable the calculation of macroscopic observables and thermodynamic transport coefficients. Understanding how curvature, torsion, and strain cooperatively dictate these statistical quantities is essential for the design of next-generation electromechanical and spintronic devices. Finally, investigating these exact strained geometries through the non-relativistic da Costa approach could bridge the theoretical gap between massless Dirac materials (like graphene) and conventional 2D electron gases (2DEGs) subjected to analogous topological deformations.
\section*{Acknowledgments}
\section*{Acknowledgments}
This work was partially supported by the Coordenação de Aperfeiçoamento de Pessoal de Nível Superior - Brasil (CAPES) - Finance Code 001, grant n. 88887.289165/2026-00 for G. M. Delgado, and Conselho Nacional de Desenvolvimento Científico e Tecnológico -CNPq grant n. 310560/2025-0 (JEGsilva).
\makeatletter
\makeatother
\end{document}